\documentclass[reprint,amsmath,amssymb,aps,superscriptaddress,nofootinbib,floatfix]{revtex4-2}

\usepackage{bm}
\usepackage{dcolumn}
\usepackage{amsfonts}
\usepackage{color}
\usepackage{braket}
\usepackage{hyperref}
\usepackage{graphicx}
\usepackage[caption=false]{subfig}

\begin{document}
\title{Symmetry-restored Skyrme-Random-Phase-Approximation calculations of the monopole strength in deformed nuclei}

\author{A. Porro}
\email{aporro@theorie.ikp.physik.tu-darmstadt.de}
\affiliation{IRFU, CEA, Universit\'e Paris-Saclay, 91191 Gif-sur-Yvette, France}
\affiliation{Technische Universit\"at Darmstadt, Department of Physics, 64289 Darmstadt, Germany}
\affiliation{ExtreMe Matter Institute EMMI, GSI Helmholtzzentrum f\"ur Schwerionenforschung GmbH, 64291 Darmstadt, Germany}

\author{G. Col\`o}
\email{gianluca.colo@mi.infn.it}
\affiliation{Dipartimento di Fisica Aldo Pontremoli, Universit\`{a} degli Studi di Milano, Via Celoria 16, 20133 Milano, Italy}
\affiliation{INFN, Sezione di Milano, Via Celoria 16, 20133 Milano, Italy}

\author{T. Duguet}
\email{thomas.duguet@cea.fr}
\affiliation{IRFU, CEA, Universit\'e Paris-Saclay, 91191 Gif-sur-Yvette, France}
\affiliation{KU Leuven, Department of Physics and Astronomy, Instituut voor Kern- en Stralingsfysica, 3001 Leuven, Belgium}

\author{D. Gambacurta}
\email{gambacurta@lns.infn.it}
\affiliation{LNS-INFN, 95123 Catania, Italy}

\author{V. Som\`a}
\email{vittorio.soma@cea.fr}
\affiliation{IRFU, CEA, Universit\'e Paris-Saclay, 91191 Gif-sur-Yvette, France}

\date{\today}

\begin{abstract}
\textbf{Background}: Within the Energy Density Functional (EDF) approach, the use of mean-field wave-functions deliberately breaking (some) symmetries of the underlying Hamiltonian is an efficient and largely utilized way to incorporate static correlations. However, the restoration of broken symmetries is eventually mandatory to recover the corresponding quantum numbers and to achieve a more precise description of nuclear properties.

\textbf{Purpose:} While symmetry-restored calculations are routinely performed to study ground-state properties and low-lying excitations, similar applications to the nuclear response are essentially limited to either formal studies or to schematic models. In the present paper, the effect of angular momentum restoration on the monopole and quadrupole responses of doubly open-shell nuclei is investigated.

\textbf{Methods:} Based on deformed Skyrme-Random Phase Approximation (RPA) calculations, the exact Angular Momentum Projection (AMP) is implemented in the calculation of the multipole strength functions, thus defining a projection after variation (PAV-RPA) scheme. The method is employed for the first time in a realistic study to investigate the effect of AMP on the coupling of monopole and quadrupole modes in $^{24}$Mg resulting from its intrinsic deformation. 

\textbf{Results:} 
The monopole PAV-RPA response function shows, in addition to the Giant Resonance peaks,  
a tremendous amount of strength in the low-energy part whose properties and nature are investigated and discussed. In the quadrupole channel, the AMP leads to a suppression of all the strength but the one corresponding to the Isoscalar Giant Quadrupole Resonance.

\textbf{Conclusions:} The nature of the anomalous low-lying monopole strength is interpreted as a contamination of the excited states via the coupling to the (non-infinitesimal) rotational motion in deformed RPA phonons. Such a spurious strength was also observed in Projected Generator Coordinate Method (PGCM) calculations based on a similar PAV approach but shown to disappear in its full Variation After Projection (VAP) counterpart. While the spurious strength could be properly subtracted in the present work, this work motivates the implementation of the full VAP-RPA in the future. 
\end{abstract}

\maketitle

\section{Introduction}\label{introduction}


In the realms of the nuclear Energy Density Functional method~\cite{LNP.879,Schunck2019,Colo2020} and of \textit{ab initio} methods~\cite{Soma:2011,Signoracci:2014dia,Tichai:2018vjc,Demol:2020mzd,Novario:2020kuf,Frosini:2021tuj,Tichai:2023epe}, allowing simple wave-functions (e.g. Slater determinants, Bogoliubov vacua or a superposition of those) to break symmetries of the Hamiltonian is an efficient way to grasp so-called static correlations in open-shell systems. The latter  typically substantiate as broken $SU(2)$ (i.e. rotational) and $U(1)$ (i.e. global-gauge) symmetries associated with the conservation of total angular momentum and particle number, respectively. Still, it is mandatory to eventually restore such symmetries when questing a good approximation to the exact solution. This is typically achieved by performing angular momentum (AMP) and particle number (PNP) projections of the symmetry-breaking state. The balance between symmetry breaking and restoration is pivotal to capture the rich diversity of nuclear phenomena, offering profound insights into the complex nature of atomic nuclei. This question is at the heart of various recent developments in \textit{ab initio} many-body theory~\cite{DuSi15,DuSi16,Qiu:2018edx,Hagen22a,Frosini22a,Frosini22c}.

Albeit symmetry-restoration techniques have been employed for a long time in EDF studies of ground-state properties and low-lying spectroscopy (see for example Ref.~\cite{Sheikh21} for a recent review), the same is not true for the linear-response theory. 
Linear-response theory within the single-reference EDF scheme boils down to solving Hartree-Fock-(Bogoliubov) (HF(B)) equations for the ground state, plus (Quasiparticle) Random Phase Approximation ((Q)RPA) equations for the excited states~\cite{RS.80}, i.e. excited states correspond to small oscillations of the ground state, or, in other words, to  nuclear
vibrations or ``phonons''. 

In singly (doubly) open-shell nuclei, the HFB solution typically breaks $U(1)$ (plus $SU(2)$) symmetry(ies) in order to capture collective static correlations associated with superfluidity (plus deformation). When $U(1)$ symmetry is broken by the HFB starting point, QRPA pair transfer probabilities were shown to overestimate the exact results in the exactly solvable Richardson model~\cite{Richardson}, the discrepancies being the largest when being close to the normal-to-superfluid phase-transition~\cite{Gambacurta2012}. The shortcoming of QRPA was traced back to the inherent violation of good particle-number associated with the breaking of $U(1)$ symmetry. It is thus of interest to explicitly investigate the symmetry contamination at the (Q)RPA performed on top of a symmetry-breaking HF(B) state.

In this context, it is often stated~\cite{Thouless61a,Brown64a} that (Q)RPA \textit{per se} restores the symmetry broken by the mean-field starting point. However, this phrasing is not free from ambiguities~\cite{Lane80a} given that the (Q)RPA wave functions only appear implicitly in the formulation. The only clear statement that can be made is that the excitation induced by the generator of the symmetry group, e.g. the infinitesimal rotation induced by the components of the angular momentum operator, is a zero-energy solution of a fully self-consistent (Q)RPA calculation~\cite{Thouless61a,RS.80}. As such, the {\it infinitesimal} rotational mode\footnote{This Goldstone mode is often denoted as "spurious" because it is not a mode that can be observed in the nuclear response.} is decoupled and orthogonal to the physical vibrational excitations of interest in (Q)RPA. This interesting property does not however imply that the (implicit) wave functions of (Q)RPA excited states carry good symmetry quantum numbers, e.g. angular momentum, and are orthogonal to {\it non}-infinitesimal rotations. In order to ensure this a full and explicit symmetry restoration is necessary.

As a matter of fact, the direct diagonalization of the Hamiltonian in the space of particle-number-projected two-quasiparticle states was shown to improve considerably the description in the Richardson model~\cite{Gambacurta2012}, thus demonstrating the importance of explicitly restoring the broken symmetry. This method, which amounts to a variation after projection (VAP) Quasiparticle Tamm-Dancoff Approximation (QTDA), has however not given birth to any realistic application since. As for the (Q)RPA, even though a full VAP-(Q)RPA formalism was designed a long time ago~\cite{FedRi85}, no realistic implementation has been performed so far. In fact, even in the context of the easier projection after variation (PAV), practitioners have relied so far on the so-called ``needle'' approximation~\cite{PhysRevC.77.034317,Peru2014}, whose validity has never been verified against actual PAV-(Q)RPA calculations\footnote{For completeness, the needle approximation to the AMP PAV-RPA scheme is derived in App.~\ref{app:needle} following the pioneering work of Ref.~\cite{Villars}.}. Focusing for example on the monopole response of doubly open-shell nuclei, which is the main interest of the present work, the needle approximation to the AMP PAV-RPA actually provides a trivial result such that an exact projection is mandatory. 

The only attempt to implement an exact AMP in PAV-RPA was performed in Ref.~\cite{Erler} based on realistic chiral interactions. However, the surprising results obtained were not analyzed in detail. It is the objective of the present paper to perform such a study within the EDF framework and deliver a comprehensive analysis of the effect of AMP in PAV-RPA. An additional motivation relates to the longstanding puzzle regarding the link between the Isoscalar Giant Monopole Resonance (ISGMR) and the nuclear matter incompressibility $K_\infty$ \cite{Blaizot1980,Garg}. The analysis of the ISGMR of some nuclei leads to consistent values of $K_\infty$, while some other nuclei appear to point towards lower values. 
While this apparent incompatibility has been solved for semi-magic (i.e. spherical) nuclei~\cite{PhysRevLett.131.082501}, doubly open-shell (i.e. deformed) nuclei still pose a challenge in this respect. In deformed nuclei, indeed, the coupling of the monopole and quadrupole response is expected to further complicate the extraction of $K_\infty$. This issue was recently studied within the deformed Skyrme-(Q)RPA approach \cite{Colo2020,Nesterenko2021,Nesterenko2022}. This leads to the question of whether explicitly restoring good angular momentum in (Q)RPA calculations is necessary to deliver a meaningful comparison with experiment in view of extracting $K_\infty$.

The paper is organized as follows. In Sec.~\ref{theory}, the inclusion of AMP to formulate PAV-TDA and PAV-RPA formalisms is presented\footnote{The restoration of $U(1)$ symmetry in QRPA is left to a future study.}. In Sec.~\ref{sec:num} specific aspects of the numerical implementation are discussed and the stability of the RPA results against parameters defining the truncated one-body basis is examined. In Sec.~\ref{results}, the AMP PAV-RPA results are presented and discussed in detail.  Eventually, conclusions are drawn in Sec.~\ref{conclu}. 

\section{Formalism}\label{theory}

\subsection{Angular Momentum Projection}
Due to the rotational invariance of the nuclear Hamiltonian, physical states $\ket{JM}$ carry good total angular momentum $J$ and angular momentum projection $M$ as quantum numbers. The remaining quantum numbers necessary to fully characterize a given quantum state are presently omitted for notation’s simplicity. Given an arbitrary state $| \Psi \rangle$ possibly breaking rotational symmetry, a state $\ket{JM}$ can be obtained by acting on it with the projection operator~\cite{RS.80}
\begin{equation}\label{eq:P}
P^J_{MK} \equiv \frac{2J+1}{8\pi^2} \int d\Omega\ {\cal D}^{J*}_{MK}(\Omega){\cal R}(\Omega)\,,
\end{equation}
where $\mathcal{D}^{J}_{MK}(\Omega)$ denotes a Wigner matrix~\cite{VMK} and $\mathcal{R}(\Omega)\equiv~e^{-i\alpha J_z}e^{-i\beta J_y}e^{-i\gamma J_z}$ is the rotation operator in three-dimensional space parameterized by the three Euler angles $\Omega\equiv(\alpha, \beta, \gamma)$\footnote{The angular momentum $J$ is presently assumed to be integer, such that the integration domain $\{\alpha\in[0,2\pi], \beta\in[0,\pi], \gamma\in[0,2\pi]\}$ and the normalising factor $1/8\pi^2$ are employed. Half-integer values of $J$ would demand to modify the $\gamma$ integration domain to $\gamma\in[0,4\pi]$ and, consequently, the normalising constant to $1/16\pi^2$.}. 

In the present study, all states to be projected are taken to remain eigenstates of $J_z$ with eigenvalue $K$. This is a consequence of retaining the axial symmetry along the $z$-axis. This simplification allows one to use the reduced form of the projection operator 
\begin{equation}
P^J_{MK} = \frac{2J+1}{2} \int_{-1}^{+1} d\left(\cos\beta \right)\ d^J_{MK}(\beta)e^{-i\beta J_y}
\end{equation}
throughout, where $\beta$ is the angle around the $y$-axis and where $d^J_{MK}(\beta)$ denotes a reduced Wigner matrix. 

\subsection{Projected multipole strength functions}

The present work implements a PAV-RPA scheme, i.e. based on the solutions of a deformed Skyrme RPA calculation, the angular momentum operator is introduced a posteriori in the computation of the multipole strength functions\footnote{As opposed to the full-fledged VAP-RPA formalism~\cite{FedRi85}, the projection operator does not enter the computation of the RPA matrix. Consequently, the present computation is free from spuriosities that forbid to use projection operators in conjunction with standard EDF parameterizations~\cite{Dobaczewski:2007ch,Lacroix:2008rj,Bender:2008rn,Duguet:2008rr,Satula:2014nba}.}. In such a scheme, the position of the peaks (i.e. the excitation energies) in the strength function are not impacted by the symmetry restoration whereas their height (i.e. the transition probabilities) can be modified.

For a given multipole operator $T_{\lambda\mu}$, the PAV-RPA strength function thus requires the computation of transition amplitudes, i.e. matrix elements, of the form
\begin{equation}
\label{eq:start1}
\braket{J_1 M_1 | T_{\lambda\mu} | J_2 M_2 }= N_1N_2\braket{\Psi_1 | P^{J_1\dagger}_{M_1K_1}T_{\lambda\mu}P^{J_2}_{M_2K_2}| \Psi_2 } \,,
\end{equation} 
with the normalization constants given by
\begin{equation}
\label{eq:start2}
N_i\equiv \left[
\langle \Psi_i|P^{J_i}_{K_iK_i}|\Psi_i\rangle
\right]^{-1/2}\,.
\end{equation} 
Employing angular momentum algebra and Wigner-Eckart's theorem, the {\it reduced} matrix element can eventually be written as~\cite{RS.80}
\begin{widetext}
\begin{align}
\langle J_1 \vert \vert T_\lambda \vert \vert J_2 \rangle =& (2J_1+1)N_1N_2\sum_{\mu=-\lambda}^{+\lambda} (-)^{J_1-K_1} 
    \begin{pmatrix}
    J_1&\lambda&J_2\\
    \text{-}K_1&\mu&K_1-\mu
    \end{pmatrix}
    \langle \Psi_1|T_{\lambda\mu}P^{J_2}_{K_1-\mu,K_2}|\Psi_2 \rangle  \label{eq:start} \\
= & \frac{(2J_1+1)(2J_2+1)}{2} N_1N_2 \sum_{\mu=-\lambda}^{+\lambda}
(-)^{J_1-K_1} \left( \begin{array}{ccc} J_1 & \lambda & J_2 \\ \text{-}K_1 & \mu & K_1-\mu \end{array}
\right)  \int_{-1}^1 d(\cos\beta)\ d^{J_2}_{K_1\text{-}\mu,K_2}(\beta) \langle \Psi_1 \vert T_{\lambda\mu} 
e^{i\beta J_y} \vert \Psi_2 \rangle \,, \nonumber
\end{align} 
\end{widetext}
with
\begin{equation}\label{eq:norm}
N_i = \left[ \frac{2J_i+1}{2} \int_{-1}^1 d(\cos\beta)d^{J_i}_{K_i,K_i}(\beta) 
\langle \Psi_i \vert e^{i\beta J_y} \vert \Psi_i \rangle \right]^{-1/2}. 
\end{equation}
In the present application, one of the two involved states, e.g. $\vert \Psi_1 \rangle$, is the symmetry-breaking ground state of the system carrying $K^\pi=K_0^{\pi_0}$, where $\pi$ denotes the parity. Whenever focusing on even-even nuclei, as is done in the following application, one has $K_0^{\pi_0}=0^+$. The final state $\vert \Psi_2 \rangle$ is an excited state carrying a given $K^\pi$.

\subsubsection{Tamm-Dancoff Approximation}

Within the TDA the ground state coincides with the HF state, the HF one-body basis being represented by the set of operators $\{a^{\dagger}_\alpha,a_\alpha\}$. This basis separates into hole (occupied) and particle (unoccupied) states. A TDA excited state $\vert n \rangle$ results from the application of a phonon operator acting on the HF ground state via a linear combination of one-particle/one-hole (ph) excitations
\begin{align}
\label{eq:TDA_1}
\vert n \rangle&\equiv Q^\dagger_n \vert {\rm HF} \rangle\nonumber\\
&\equiv \sum_{ph} X^{ph}_n a^\dagger_p a_h \vert \text{HF} \rangle\, .
\end{align} 
In this case, the matrix element entering Eq.~(\ref{eq:start}) reads explicitly as
\begin{align}
\langle {\rm HF} \vert T_{\lambda\mu} P^{J}_{K_0\text{-}\mu, K} \vert n \rangle  &= \langle {\rm HF} \vert T_{\lambda\mu} P^{J}_{K_0\text{-}\mu, K} 
Q^\dagger_n \vert {\rm HF} \rangle \label{eq:TDA_2} \\
&= \sum_{ph} X^{ph}_n \langle {\rm HF} \vert T_{\lambda\mu} P^{J}_{K_0\text{-}\mu, K} a^\dagger_p a_h \vert {\rm HF} \rangle\,.\nonumber
\end{align} 

\subsubsection{Random Phase Approximation}

Within the RPA, $\vert \Psi_1 \rangle$ is the correlated RPA ground state, hereafter indicated as $\vert {\rm RPA} \rangle$. A RPA excited state $\vert n \rangle$ reads now as
\begin{align}
\vert n \rangle&\equiv Q^\dagger_n \vert {\rm RPA} \rangle\nonumber\\
&\equiv \sum_{ph} \left( X^{ph}_n a^\dagger_p a_h \vert {\rm RPA} \rangle - 
Y^{ph}_n a^\dagger_h a_p \vert {\rm RPA} \rangle \right)\,,
\end{align}
where the phonon operator now also includes a de-excitation component associated with the  $Y$-amplitudes. The correlated RPA ground state satisfies the vacuum condition $Q_n \vert {\rm RPA} \rangle = 0$, for all $n$. 

The na\"{\i}ve application of AMP that could work for the TDA leads here to the vanishing of the contribution from the $Y$-amplitudes. The PAV-RPA transition amplitudes must rather be derived on the basis of the quasi-boson approximation (QBA) that is invoked to obtain standard RPA equations. Such an approximation boils down to replacing operator products with appropriate commutators and the RPA ground-state with the HF ground state in a second step\footnote{The two steps do not commute.}. In the simple RPA case, the procedure reads as\footnote{Here, and in the following, equalities in the QBA sense are indicated by wiggly equal signs.}
\begin{align}
	\label{eq:standardRPA}
	\langle n \vert T_{\lambda\mu} \vert {\rm RPA} \rangle&= \langle {\rm RPA} \vert T_{\lambda\mu} Q^\dagger_n \vert {\rm RPA} \rangle\nonumber\\
	&=\langle {\rm RPA} \vert \left[ T_{\lambda\mu},  Q^\dagger_n \right] \vert {\rm RPA} \rangle \nonumber\\
	&\approx \langle {\rm HF} \vert \left[ T_{\lambda\mu},  Q^\dagger_n \right] \vert {\rm HF} \rangle\nonumber\\
	&=\sum_{ph}\Bigl\{X^{ph}_n\braket{h|T_{\lambda\mu}|p}+Y^{ph}_n\braket{p|T_{\lambda\mu}|h}\Bigr\}, 
\end{align}
where the vacuum character of the RPA ground state was used in the second equality to add a null term. Notice that the replacement of the RPA ground state with the HF one in Eq.~\eqref{eq:standardRPA} does not amount to a neglect of ground-state correlation, but is instead coherent with the quasiboson approximation~\cite{Stetcu02b}. With the goals of avoiding the suppression of the backward amplitudes $Y$ in presence of the symmetry projector and recovering Eq.~(\ref{eq:standardRPA}) in absence of it, one can write in close similarity to the RPA case 
\begin{widetext}
\begin{align}\label{eq:RPA_wrong}
\langle {\rm RPA} \vert T_{\lambda\mu} P^{J}_{K_0-\mu, K} \vert n \rangle  &= \langle {\rm RPA} \vert T_{\lambda\mu} P^{J}_{K_0-\mu, K} Q^\dagger_n \vert {\rm RPA} \rangle \nonumber\\
&= \langle {\rm RPA} \vert \left[ T_{\lambda\mu} P^{J}_{K_0-\mu, K},  
Q^\dagger_n \right] \vert {\rm RPA} \rangle\nonumber \\
& \approx  \langle {\rm HF} \vert T_{\lambda\mu} P^{J}_{K_0-\mu, K} Q^\dagger_n - Q^\dagger_n  T_{\lambda\mu} P^{J}_{K_0-\mu, K} \vert {\rm HF} \rangle \nonumber \\
& =   \sum_{ph} X^{ph}_n \langle {\rm HF} \vert T_{\lambda\mu} P^{J}_{K_0-\mu, K} a^\dagger_p a_h \vert {\rm HF} \rangle
+ Y^{ph}_n \langle {\rm HF} \vert a^\dagger_h a_p T_{\lambda\mu} P^{J}_{K_0-\mu, K} \vert {\rm HF} \rangle\, .
\end{align}
However, the contribution from the backward amplitudes cancels out for all $K\neq K_0$ or $\mu\neq0$, which corresponds to artificially restrictive selection rules. It is rather preferable to write
\begin{align}\label{eq:RPA_right}
\langle {\rm RPA} \vert T_{\lambda\mu} P^{J}_{K_0-\mu, K} \vert n \rangle  & =  \langle {\rm RPA} \vert T_{\lambda\mu} P^{J}_{K_0-\mu, K} 
Q^\dagger_n \vert {\rm RPA} \rangle\nonumber\\
&=\langle {\rm RPA} \vert T_{\lambda\mu} P^{J}_{K_0-\mu, K} 
Q^\dagger_n - Q^\dagger_n P^{J}_{K_0-\mu, K}  T_{\lambda\mu}   \vert {\rm RPA} \rangle\nonumber \\
& \approx  \langle {\rm HF} \vert T_{\lambda\mu} P^{J}_{K_0-\mu, K} Q^\dagger_n - Q^\dagger_n P^{J}_{K_0-\mu, K}  T_{\lambda\mu}  
\vert {\rm HF} \rangle\nonumber \\
& =   \sum_{ph} X^{ph}_n \langle {\rm HF} \vert T_{\lambda\mu} P^{J}_{K_0-\mu, K} a^\dagger_p a_h \vert {\rm HF} \rangle
+ Y^{ph}_n \langle {\rm HF} \vert a^\dagger_h a_p P^{J}_{K_0-\mu, K} T_{\lambda\mu} \vert {\rm HF} \rangle.
\end{align}
The term added after the second equality, with $Q^\dagger_n$ acting on the RPA bra state, is null as was the term\footnote{The two terms differ by the arbitrary ordering of $T_{\lambda\mu}$ and $P^{J}_{K_0-\mu, K}$.} added by the commutator in the second equality of Eq.~\eqref{eq:RPA_wrong}. Equation~\eqref{eq:RPA_right} now delivers a non-vanishing contribution from the $Y$ amplitudes for all $K$. Furthermore, the result correctly reduces to the TDA one when ignoring backward amplitudes and to the original RPA transition amplitudes when removing the projector. Inserting Eq.~\eqref{eq:RPA_right} into Eq.~\eqref{eq:start}, one eventually obtains the PAV-RPA reduced transition matrix element under the form
\begin{eqnarray}
\label{eq:final}
\langle {\rm RPA} \vert \vert T_\lambda \vert \vert n \rangle & = & (2J_0+1) N_0N_n 
(-1)^{J_0-K_0} \sum_{ph}
\sum_\mu \left[ X_n^{ph} + (-1)^\mu Y_n^{ph} \right] 
\left( \begin{array}{ccc} J_0 & \lambda & J \\ -K_0 & \mu & K_0-\mu \end{array}
\right) \nonumber \\
&& \times \int_{-1}^1 d(\cos\beta)\ d^{J}_{K_0-\mu,K}(\beta) \langle {\rm HF} \vert T_{\lambda\mu} 
e^{i\beta J_y} a^\dagger_p a_h \vert {\rm HF} \rangle \, .
\end{eqnarray} 
The normalization factors from Eq.~\eqref{eq:norm} are derived consistently and read as
\begin{equation}\label{eq:n0}
N_0 = \left[ \int_{-1}^1 d(\cos\beta)\ d^{J_0}_{K_0,K_0}(\beta) 
\langle {\rm HF} \vert e^{i\beta J_y} \vert {\rm HF} \rangle \right]^{-1/2} \, , 
\end{equation}
and
\begin{equation}\label{eq:nn}
N_n = \left[ \sum_{php'h'} \left( X_n^{ph}X_n^{p'h'} - Y_n^{ph}Y_n^{p'h'} \right)
\int_{-1}^1 d(\cos\beta)\ d^{J}_{K,K}(\beta) 
\langle {\rm HF} \vert a^\dagger_{h'}a_{p'} e^{i\beta J_y} a^\dagger_p a_h \vert {\rm HF} \rangle \right]^{-1/2} \, .
\end{equation}
\end{widetext}
Notice that the ground-state normalization $N_0$ in Eq.~\eqref{eq:n0} deliberately omits ground-state correlation. This modifies the multipole response by an overall multiplicative factor and will, thus, not affect the main conclusions of the present study. The inclusion of ground-state correlations in this ground-state normalisation is in fact non-trivial, as clearly shown in Ref.~\cite{Johnson02a,Stetcu02a}, and goes beyond the aim of the present work.

\subsection{Rotated RPA state}

As a relevant element for the following discussion, the RPA ground-state projected on angular momentum $J=0$ is considered. Given that $d^0_{00}(\beta)=1$, this state corresponds to the equally weighted superposition of the deformed RPA state rotated in all possible orientations in space and is thus denoted below as the {\it rotated} RPA state 
\begin{align}
\label{eq:def_ROT}
\ket{\text{ROT}}&\equiv P^0_{00} \vert {\rm RPA} \rangle \nonumber\\
&=\frac{1}{2} \int_{-1}^{+1} d\left( \cos\beta \right)\ e^{-i\beta J_y} \vert {\rm RPA} \rangle\,.
\end{align}
The overlap with RPA excited states is eventually introduced as
\begin{align}
	\langle {\rm ROT} \vert n \rangle&\equiv\langle {\rm RPA} \vert P^0_{00} \vert n \rangle\nonumber\\
	&\approx\langle {\rm HF} \vert P^0_{00} \vert n \rangle\,,
	\label{eq:overlap}
\end{align}
where the second equality relates to the QBA.

\section{Numerical details}
\label{sec:num}

\begin{figure*}
	\subfloat[\label{fig:conv_RPA_a}]{\includegraphics[width=0.4\textwidth]{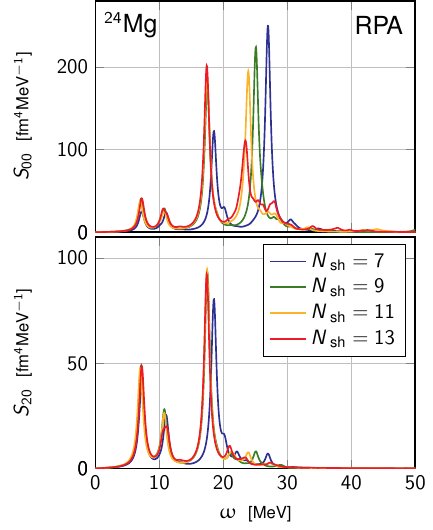}}
    \hspace{0.5cm}
	\subfloat[\label{fig:conv_RPA_b}]{\includegraphics[width=0.4\textwidth]{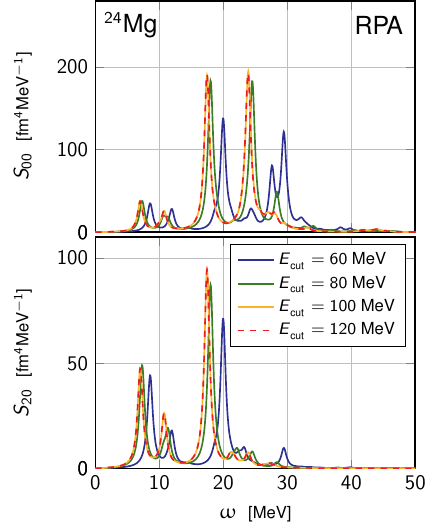}}
	\caption{\textbf{(a)} monopole (upper panel) and quadrupole (lower panel) RPA responses in  $^{24}$Mg for different $N_{\rm sh}$ values ($E_{\rm cut}=~100$~MeV). \textbf{(b)} same but for different $E_{\rm cut}$ values ($N_{\rm sh}=~11$).}
\end{figure*}

Results are presented in this work for $^{24}$Mg, using the SkM$^*$ Skyrme EDF~\cite{Bartel1982}. The QRPA code originally introduced in Ref.~\cite{Losa} is used.
This is based on the HFB solution in axial symmetry delivered by the HFBTHO code~\cite{Stoitsov05a} using a spherical harmonic oscillator (sHO) basis. 
Without AMP, this numerical scheme was employed in Ref.~\cite{Colo2020-1} to compute  monopole and quadrupole strengths in Molybdenum isotopes. 
The sHO basis is characterized by the value of $\hbar\omega$=~1.2$\ \times \ 41/{\rm A}^{1/3}$ [MeV] and by a number of major oscillator shells $N_{\rm sh}$.

The axial quadrupole deformation parameter defined as~\cite{Stoitsov05a}
\begin{equation}
	\beta\equiv\sqrt{\frac{\pi}{5}}\frac{\braket{Q_{20}}_\pi+\braket{Q_{20}}_\nu}{\braket{r^2}_\pi+\braket{r^2}_\nu}\,,
\end{equation}
is employed to characterize the reference state that is presently found to be a HF (i.e. non superfluid) solution for all considered basis size values $N_{\rm sh}$. Characteristics of this HF solution and their dependence on $N_{\rm sh}$ can be found in Tab.~\ref{tab:HF_prpa_conv}.
\setlength\tabcolsep{10pt}
\begin{table}
	\centering
	\begin{tabular}{cccc}
		\hline
		\hline
		$N_{\rm sh}$  & $E_{\text{HF}}$ [MeV] & $r$ [fm] & $\beta$ \\ 
		\hline
		7 & -195.65 & 2.991 & 0.378 \\ 
		9 & -196.21 & 3.009 & 0.392 \\ 
		11 & -196.93 & 3.011 & 0.383 \\ 
		13 & -197.15 & 3.016 & 0.390 \\ 
		\hline
		\hline
	\end{tabular}
	\caption{$^{24}$Mg Hartree-Fock energy $E_{\text{HF}}$, root-mean-square radius $r$ and axial quadrupole deformation $\beta$ as a function of the HO basis-size parameter $N_{\rm sh}$.}
	\label{tab:HF_prpa_conv}
\end{table}

The deformed QRPA problem, which here reduces to deformed RPA, is solved in matrix form
\begin{equation}
	\label{RPA} 
	\left( \begin{array}{cc}
		A & B \\
		-B^* &- A^* \end{array}  \right) \left( \begin{array}{c}
		X^{n} \\
		Y^{n}  \end{array} \right) =E_{n} \left( \begin{array}{c}
		X^{n} \\
		Y^{n}  \end{array} \right)\, ,
\end{equation}
by using diagonalization techniques for sparse matrices. In the present work, RPA equations are solved for $K^\pi=0^+$ and the strength associated with standard isoscalar monopole and quadrupole operators, i.e. $\sum_i r_i^2$ and $\sum_i r_i^2 Y_{20}$ respectively, are computed. In all figures, discrete strengths are averaged using Lorentzian functions with a width of $\Gamma=1.0$~MeV.

The stability of RPA strength functions is displayed in Fig.~\ref{fig:conv_RPA_a} against the sHO basis-size parameter $N_{\rm sh}$ and in Fig.~\ref{fig:conv_RPA_b} against the parameter $E_{\rm cut}$ corresponding to the maximum energy of $K^\pi=0^+$ p-h excitations included in the calculation. 
Fixing $E_{\rm cut}=100$~MeV, the quadrupole RPA response displays a converging pattern as a function of  $N_{\rm sh}$ in the bottom panel of Fig.~\ref{fig:conv_RPA_a}. The low-energy components (below $\approx$ 20~MeV) of the monopole response (upper panel) are also seen to converge for relatively small $N_{\rm sh}$. However, the position of the higher-energy peak  shows a strong dependence on  $N_{\rm sh}$ and its fragmentation is still increasing for the largest model-space employed. This phenomenon is attributed to high-lying excitations involving states in the continuum, such that details of single-particle configurations strongly affect the global response.
Fixing now the number of shells in the sHO basis to $N_{\rm sh}=11$, monopole and quadrupole RPA responses are shown to be essentially identical for $E_{\rm cut}=$100~MeV and $E_{\rm cut}=$120~MeV in Fig.~\ref{fig:conv_RPA_b}. Unless otherwise specified, results displayed below are obtained using $N_{\rm sh}=11$ and $E_{\rm cut}=$100~MeV. 

In order to obtain the PAV-RPA strengths of interest, Eqs.~\eqref{eq:final}, \eqref{eq:n0} and \eqref{eq:nn} have been implemented to accommodate the RPA solutions discussed above. Such an implementation, currently limited to $K=0$, is based on a series of basis transformations building on the matrix elements of the operator $J_y$ computed in the sHO basis. Details can be found in Ref.~\cite{Porro2023}. The implementation was validated through several steps, including the application to a spherical system, i.e. $^4$He, where the PAV-RPA and  the original RPA strengths coincide up to numerical precision.

\section{Results and discussion}
\label{results}

\begin{figure}[b]
	\includegraphics[width=0.4\textwidth]{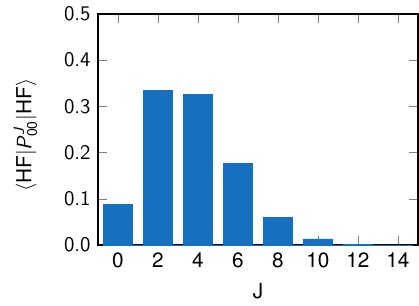}
	\caption{Angular momentum decomposition of the HF ground state in $^{24}$Mg ($N_{\rm sh}=~11$)}.\label{fig:HF_J}
\end{figure}

\subsection{Symmetry breaking of the Hartree-Fock state}

The angular-momentum decomposition of the HF ground state is displayed in Fig.~\ref{fig:HF_J} and is similar to the one obtained in Ref.~\cite{Marevic2022}. Consistently with the rather large quadrupole deformation found for the HF minimum ($\beta=0.38$), the HF wave function spreads over several (even) $J$ values, the dominant components being found for $J = 2, 4$.

\begin{figure*}
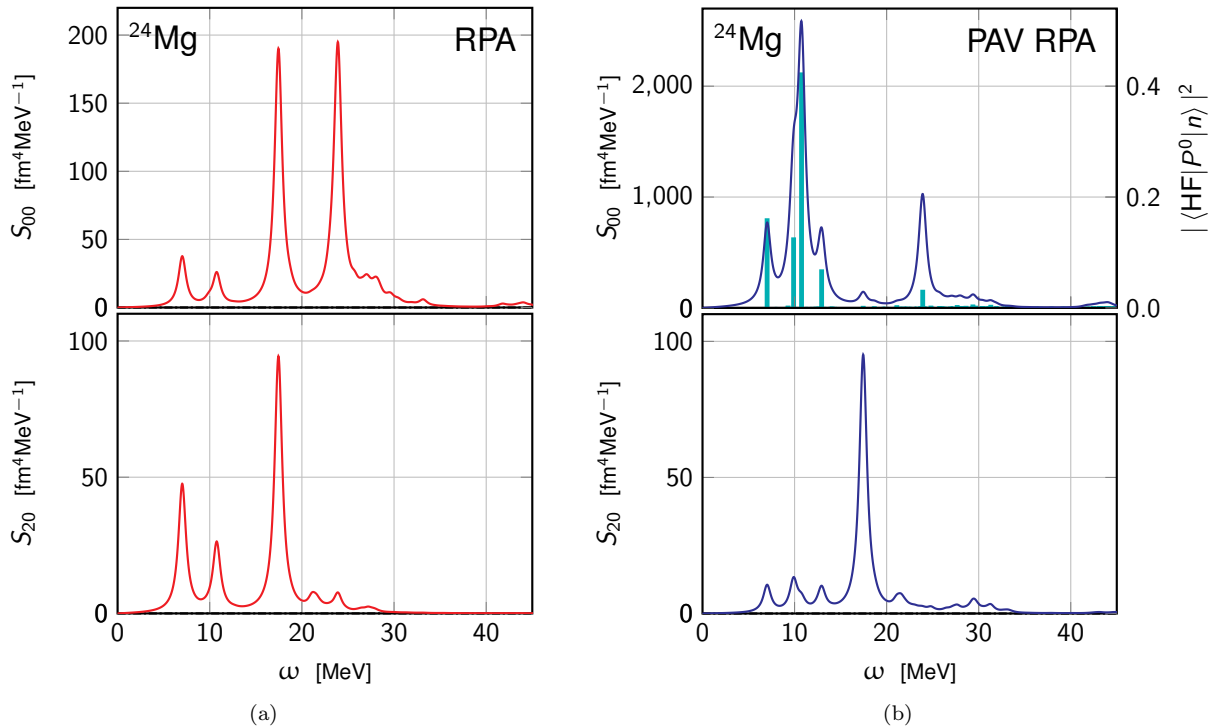

\subfloat[\label{fig3a}]{\includegraphics[width=0.4\textwidth]{Mg24\_RPA\_IS.pdf}}
\hspace{0.5cm}
\subfloat[\label{fig3b}]{\includegraphics[width=0.473\textwidth]{Mg24\_RPA\_IS_proj.pdf}}
\caption{\textbf{(a)} monopole (upper panel) and quadrupole (lower panel) RPA responses in $^{24}$Mg ($N_{\rm sh}=$11 and $E_{\rm cut}=100$~MeV). \textbf{(b)} same for PAV-RPA responses. The upper-right panel additionally contains the overlap between RPA excited states and the rotated ground state (Eq.~\eqref{eq:overlap}) up to a normalization factor.}
\end{figure*}

\subsection{RPA versus PAV-RPA strengths}

Based on such a starting point, RPA and PAV-RPA strength functions are compared respectively in Figs.~\ref{fig3a} and \ref{fig3b}, with the monopole response in the upper panels and the quadrupole response in the lower panels. At 17.5~MeV, the Isoscalar Giant Quadrupole Resonance (ISGQR) stands out in the quadrupole RPA strength. This peak aligns closely with the low-energy component of the ISGMR that additionally displays a higher-energy component at 24~MeV. While the latter is the reminiscence of the ISGMR in spherical nuclei, the former results from the coupling to the $K=0$ component of the ISGQR. Furthermore, a substantial amount of low-energy strength below 15 MeV is also found in both channels, that has no obvious interpretation. 

Including the AMP, the ISGQR remains unchanged whereas the low-energy part of the quadrupole strength is  suppressed. The monopole spectrum is much more substantially altered  (notice the the scale on the y-axis) by the symmetry restoration.
Indeed, the low-energy strength is strongly enhanced and dominates over the ISGMR peaks whose high-energy component is also substantially increased. A similar behaviour was observed in other nuclei, e.g. $^{20}$Ne and $^{28}$Si~\cite{Porro2023}.

\subsection{Spurious rotational coupling}

The impact of the angular momentum projection on the monopole strength seems largely anomalous. It was thoroughly checked that this feature is not a numerical artefact and is stable with respect to $N_{\rm sh}$ and $E_{\rm cut}$. Similar results were already obtained in Ref.~\cite{Erler} but no explanation was given. It is thus necessary to shed light on the nature of the anomalous low-energy strength resulting from the symmetry restoration. 

In the upper panel of Fig.~\ref{fig3b}, the overlap (Eq.~\eqref{eq:overlap}) between each excited RPA state and the rotated RPA ground state is displayed, up to a normalizing factor, as (blue) bars in connection to the right axis.
The overlap is large for states in the low-energy region, exactly where the monopole strength was anomalously enhanced by the angular-momentum projection. Large overlaps indicate that these deformed RPA excited states are not strictly vibrational but rather display a significant coupling to a rotational motion of the nucleus. As reminded in the introduction, the infinitesimal rotational motion present in the $K^\pi=1^+$ channel\footnote{The quantum numbers of the infinitesimal rotation generated by the rotation operator $e^{i\beta J_y}$ at lowest order in $\beta$, i.e. by the linear term in $J_y$, is indeed $K^\pi=1^+$.} is properly decoupled from actual vibrational excitations in RPA. However, the present analysis demonstrates that it is not the case for non-infinitesimal rotation that can furthermore pollute different $K^\pi$ channels (see discussion on p.~145 of Ref.~\cite{BM2}). 

\begin{figure*}
    \subfloat[\label{fig:dens_a}]{\includegraphics[width=0.4\textwidth]{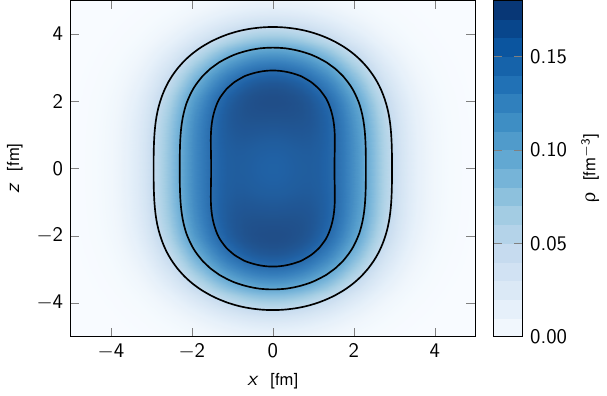}}\hspace{1cm}
    \subfloat[\label{fig:dens_b}]{\includegraphics[width=0.4\textwidth]{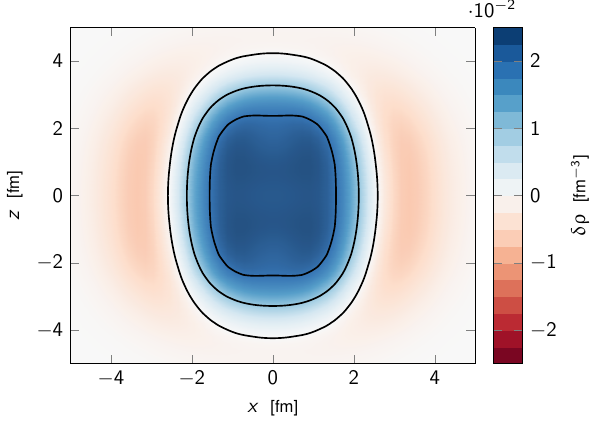}}\hfill
    \subfloat[\label{fig:dens_c}]{\includegraphics[width=0.4\textwidth]{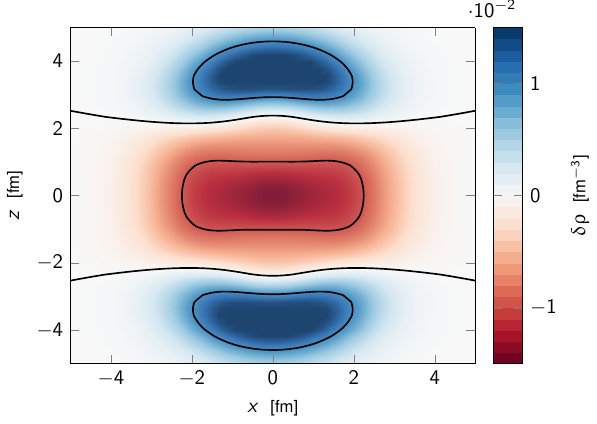}}\hspace{1cm}
    \subfloat[\label{fig:dens_d}]{\includegraphics[width=0.4\textwidth]{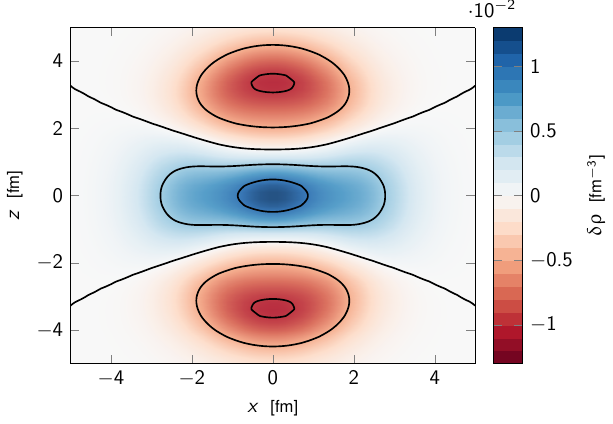}}
	\caption{$^{24}$Mg \textbf{(a)} one-body intrinsic matter density of the HF ground state, \textbf{(b)} intrinsic RPA transition matter density to the ISGMR phonon at 23.9~MeV, \textbf{(c)} intrinsic RPA transition matter density to the ISGQR phonon at 17.5~MeV and \textbf{(d)} intrinsic RPA transition matter density to the phonon at 10.8~MeV.}
	\label{fig:Mg24_dens}
\end{figure*}

In order to have a complementary view of the rotational character of deformed RPA phonons, as well as a confirmation of the expected character of the ISGMR and ISGQR, transition densities $\delta\rho$ associated with several excited states of interest are displayed in Fig.~\ref{fig:Mg24_dens}. A representation in cylindrical coordinates is employed, so that the vertical axis is the $z$-axis and the horizontal axis is one of the possible equivalent axes in the perpendicular plane. 
For reference, the HF ground-state density $\rho_0$ is shown in Fig.~\ref{fig:dens_a}, displaying the typical shape of a quadrupole deformed system. 
In Fig.~\ref{fig:dens_b}, the transition density to the main peak of the ISGMR at 23.9~MeV is shown, indeed corresponding to a typical monopole oscillation of the ground-state shape. 
The transition density to the ISGQR peak at 17.5~MeV shown in Fig.~\ref{fig:dens_c} displays a nodal line at constant polar angle $\pm \theta$ that is typical of an excitation induced by $Y_{20} \propto \left( 3\cos^2 \theta - 1 \right)$. 
\begin{figure*}[t]
\subfloat[\label{fig4a}]{\includegraphics[width=0.4\textwidth]{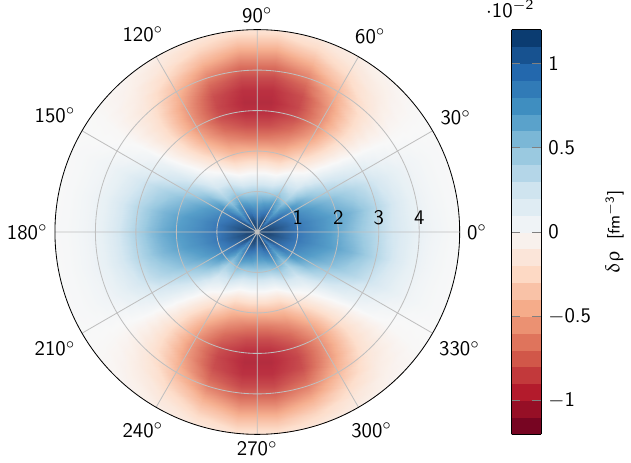}}
\hspace{0.5cm}
\subfloat[\label{fig4b}]{\includegraphics[width=0.4\textwidth]{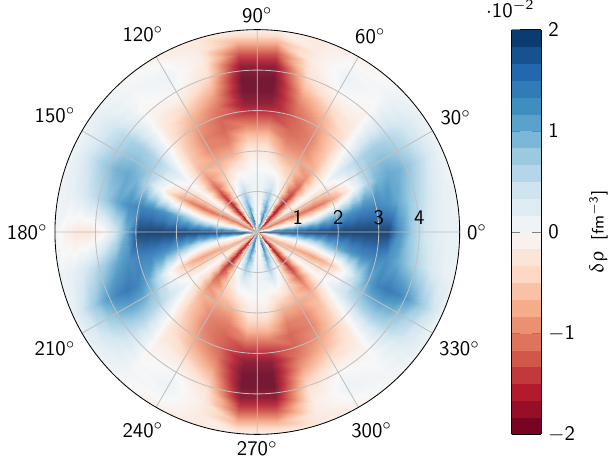}}
\caption{\textbf{(a)} polar coordinate representation of the intrinsic RPA transition density to the rotational phonon at 10.8~MeV in $^{24}$Mg. \textbf{(b)} same for the combination of two densities obtained by rotating the HF state by $\beta\approx \pm 24^\circ$. }
\end{figure*}
Finally, Fig.~\ref{fig:dens_d} displays the transition density to the low-energy peak at about 10~MeV carrying the largest strength in the monopole PAV-RPA response. 

Based on this picture, the nature of the state remains elusive. Therefore, a polar coordinate representation was further adopted to make easier the rotational character of the state. In Fig.~\ref{fig4a}, the same transition density is displayed as a function of $r$ and $\theta$. 
It is found that the overlap between the wave function of the low-energy peak at 10.8~MeV and the rotated RPA ground state is maximal when the rotation angle is $\beta\approx \pm 24^\circ$. Thus, Fig.~\ref{fig4b} displays the arithmetic average of the two densities obtained by rotating the HF ground state by $\beta\approx \pm 24^\circ$. The comparison clearly shows that, in the surface region (i.e., $r\approx$ 3~fm), the two densities have the same maxima, minima, and nodal points, thus confirming the strong rotational character of the RPA phonon. 

\subsection{Subtracted PAV-RPA strengths}

While hindered in the unprojected RPA strength function, the significant rotational component of low-energy RPA phonons is strongly enhanced in the monopole response when restoring good angular momentum. 
Even though the high component of the ISGMR is also enhanced, albeit much less than the low-energy peaks, the ISGQR and the associated peak of the ISGMR are essentially unaffected due to their negligible rotational content. Eventually, the anomalous rotational component of low-energy phonons hinders the physical information of interest in the PAV-RPA strength functions. Consequently, a procedure to subtract it is now formulated and implemented.

Given an excited RPA state $\vert n \rangle$, a new state is defined by subtracting its rotational component according to
\begin{equation}
\vert \tilde n \rangle \equiv N_{\tilde{n}} \left( \vert n \rangle - a_n \vert {\rm ROT} \rangle \right)
\end{equation}
with $N_{\tilde{n}}$ a normalization factor. The constant $a_n$ is chosen to make $\vert \tilde n \rangle$  orthogonal to $\vert {\rm ROT} \rangle$, i.e. to ensure
\begin{equation}
	\label{eq:orthogonal}
	\langle {\rm ROT} \vert \tilde n \rangle = 0\,,
\end{equation}
which leads to
\begin{equation}
a_n =  \frac{\langle {\rm ROT} \vert n \rangle}{\langle {\rm ROT} \vert  {\rm ROT} \rangle}\,.
\label{eq:coeff}
\end{equation}
It is straightforward to check that
\begin{equation}
(N_{\tilde{n}})^{-2} = 1 - \vert a_n \vert^2\braket{\text{ROT}|\text{ROT}}\,.
\end{equation}
A similar procedure was already applied to deal with the spurious translational motion~\cite{Colo2000} and to the one associated with  number-symmetry breaking and restoration in HFB plus QRPA~\cite{PhysRevC.78.064304}. 

Subsequently, a set of projected subtracted states are introduced according to 
\begin{equation}
	\ket{\tilde{n}^{JM}}\equiv N_{\tilde{n}}^JP^J_{MK}\ket{\tilde{n}}\,.
	\label{eq:sub_PAV_RPA}
\end{equation}
Because of the definition of $\ket{\rm ROT}$ in Eq.~\eqref{eq:def_ROT}, $\ket{\tilde{n}^J}$ differs from the original PAV-RPA state only for $J = 0$, in which case the normalizing factor reads as
\begin{align}
	(N_{\tilde{n}}^0)^{-2}&=\braket{n|P^0_{00}|n}-|a_n|^2\braket{\text{ROT}|\text{ROT}}\,.
\end{align}
The value of $\braket{\text{ROT}|\text{ROT}}$ is set to fulfil the condition $(N_{\tilde{n}}^0)^{-2}=0$ for the RPA phonon that most strongly couples to the rotational state. The corresponding phonon is eventually removed from the spectrum, as is traditionally done when dealing with the spurious phonon associated with the infinitesimal rotation in the $K^\pi=1^+$ channel.

Results obtained employing the subtraction technique defined above are labelled as subtracted PAV-RPA in the following. It is worth noting that performing the AMP and subtracting the rotational state a posteriori leads to the same results; i.e. the subtraction and the projection commute. Furthermore, it was checked that subtracted PAV-RPA strength functions are stable against variations of $N_{\rm sh}$ and $E_{\rm cut}$.

\begin{figure}
\includegraphics[width=0.4\textwidth]{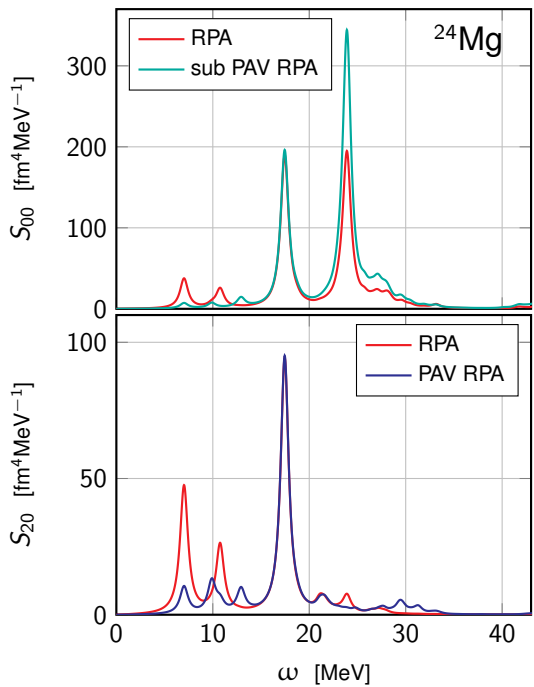}
\caption{$^{24}$Mg  monopole (upper panel) and quadrupole (lower panel) RPA and PAV-RPA responses ($N_{\rm sh}=$11 and $E_{\rm cut}=100$~MeV). In the monopole case the subtracted PAV-RPA response is also shown.}\label{fig:sub_PAV}
\end{figure}

The subtracted monopole PAV-RPA response of $^{24}$Mg is displayed in the upper panel of Fig.~\ref{fig:sub_PAV} and compared to the original RPA results.
The RPA and PAV-RPA quadrupole strengths already appearing in Figs. \ref{fig3a} and \ref{fig3b} are shown in the lower panel of Fig.~\ref{fig:sub_PAV} for reference. 
Once the rotational component has been removed, the monopole response becomes weakly affected by the AMP, similarly to the quadrupole response.
The only exception concerns the high-energy component of the ISGMR, whose strength is enhanced relative to the lower peak associated with $K=0$ component of the ISGQR. 
Eventually, the low-energy response is now meaningful thanks to the subtraction procedure and significantly suppressed by the AMP for both multipoles. 
It is worth stressing that this procedure orthogonalize the $J = 0$ components of the excited states with respect to the $J = 0$ component of the ground state. A fully orthogonal set of projected-RPA states would be achieved only in a full VAP approach.

Analogous results were recently obtained based on Projected Generator Coordinate Method (PGCM) calculations in light- and medium-mass nuclei within an \textit{ab initio} scheme~\cite{Porro2023,Porro24a}. Starting from unprojected GCM solutions, similar in essence to symmetry-breaking RPA calculations\footnote{It is worth remembering the (Q)RPA can be obtained as the harmonic limit of the GCM~\cite{jancovici64a,Porro2023}.}, the AMP is introduced  into the strength function after having solved the Hill-Wheeler GCM variational equation. This defines the PAV-GCM scheme. Doing so, a spurious coupling to the rotational motion arises as in PAV-RPA calculations, which can be similarly removed via subtraction techniques. Furthermore, differently from the present RPA frame, the full PGCM calculation enforcing the AMP while solving the variational equation, i.e. the VAP-GCM scheme, is also currently accessible. This allows one to properly handle the rotation-vibration coupling and avoid from the outset to contaminate the symmetry-restored strength functions with spurious contributions. Indeed, the PGCM monopole strengh function shows only modest differences with the unprojected GCM one and does not display the anomalous low-energy strength visible in PAV-GCM. Interestingly, full-fledged VAP-GCM (i.e. PGCM) results are close but not identical to subtracted PAV-GCM ones~\cite{Porro2023,Porro24a}. In particular, the position of the peaks (i.e. the excitation energies) is impacted by the symmetry restoration in the VAP-GCM whereas it is left unchanged by construction in any (subtracted) PAV scheme. On the RPA side, the results obtained in the present work and the complementary ones obtained via the PGCM strongly motivate the realistic implementation of the full-fledged VAP-(Q)RPA scheme~\cite{FedRi85} in the near future.

\section{Conclusions}\label{conclu}

The effects of angular momentum projection on the strength functions originating from symmetry-breaking RPA calculations have been studied in the case of the well-deformed prolate nucleus $^{24}$Mg.

The appearance of anomalously large contributions in the corresponding PAV-RPA monopole strength function at low energy was observed and attributed to a spurious coupling of deformed RPA phonons to the (non-infinitesimal) rotational motion.
A similar behaviour was also identified recently in PAV-GCM calculations~\cite{Porro2023,Porro24a}, i.e. the coupling to the rotational states related to the symmetry-breaking nature of the reference state is not peculiar to a specific many-body method used to compute vibrational excitations.

In deformed RPA, the spurious solution in the $K^\pi=~1^+$ channel associated with an infinitesimal rotation appears as a zero-energy solution and can be subtracted from the spectrum. However, RPA is not suited to separate genuine, i.e. non-infinitesimal, rotations that contaminate its eigenstates and are anomalously magnified in the monopole strength when restoring good angular momentum.


A strategy to explicitly isolate and subtract the rotational content of the RPA phonons was thus introduced and shown to restore meaningful AMP PAV-RPA strengh functions. Still, correcting the problem a posteriori is not entirely satisfactory: the proper treatment of coupling effects between rotational and vibrational motions can only be achieved if the AMP is considered while solving the (Q)RPA equations, i.e. by implementing the full-fledged AMP VAP-(Q)RPA~\cite{FedRi85} in realistic calculations. As a matter of fact, the equivalent method within the realm of the GCM, i.e. the PGCM, is shown to fully take care of this issue~\cite{Porro2023,Porro24a}. Work to parallel such PGCM calculations within the frame RPA methods is thus mandatory.
\vspace{0.5cm}
\section*{Acknowledgments}
The authors wish to thank Mikael Frosini for useful clarifications about the angular momentum projection, as well as Kenichi Yoshida and Naoyuki Itagaki for valuable discussions. A.P. was supported by the CEA NUMERICS program, which has received funding from the European Union's Horizon 2020 research and innovation program under the Marie Sk{\l}odowska-Curie grant agreement No 800945, and by the Deutsche Forschungsgemeinschaft (DFG, German Research Foundation) – Projektnummer 279384907 – SFB 1245. A.P. was also supported by ACRI - Associazione di Fondazioni e di Casse di Risparmio Spa - in the context of the Young Investigator Training program 2019. 

\appendix

\section{Derivation of the needle approximation}
\label{app:needle}

Let us start from Eqs. (\ref{eq:start}) and (\ref{eq:norm}):
\begin{widetext}
\begin{eqnarray}
\langle J_1 \vert \vert T_\lambda \vert \vert J_2 \rangle & = & \frac{(2J_1+1)(2J_2+1)}{2} N_1N_2 \sum_\mu
(-)^{J_1-K_1} \left( \begin{array}{ccc} J_1 & \lambda & J_2 \\ -K_1 & \mu & K_1-\mu \end{array}
\right) \nonumber \\
&& \times \int_{-1}^1 d(cos\beta)\ d^{J_2}_{K_1-\mu,K_2}(\beta) \langle \Phi_1 \vert T_{\lambda\mu} 
e^{i\beta J_y} \vert \Phi_2 \rangle, \nonumber
\end{eqnarray} 
with
\begin{equation}
N_i = \left[ \frac{2J_i+1}{2} \int_{-1}^1 d(cos\beta)\ d^{J_i}_{K_i,K_i}(\beta) 
\langle \Phi_i \vert e^{i\beta J_y} \vert \Phi_i \rangle \right]^{-1/2} \, , \nonumber
\end{equation}
where $e^{i\beta J_y}$ operates a rotation around one of the axis perpendicular to the symmetry
axis. If the nucleus is well deformed, a first approximation assumes that the wave function $\vert \Phi \rangle$ has zero overlap with its rotated counterpart for not too small angles. Given the property
\begin{equation}
e^{i\pi J_y} \vert \Phi_K \rangle = \vert \Phi_{-K} \rangle,
\end{equation}
rotating the state by an angle close to $\pi$ delivers a state that a strong overlap with the original state for $K=0$. 

Based on the above, the normalization factor can be written as
\begin{eqnarray}\label{eq:appr1}
\int_{-1}^1 d(cos\beta)\ d_{K_i,K_i}^{J_i}(\beta) \langle \Phi_i \vert e^{i\beta J_y} 
\vert \Phi_i \rangle & = &
\int_0^\epsilon d\beta\ sin(\beta) d_{K_i,K_i}^{J_i}(\beta) N(\beta) \nonumber \\
& + & 
\int_{\pi-\epsilon}^\pi d\beta\ sin(\beta) d_{K_i,K_i}^{J_i}(\beta) N(\beta), \nonumber \\
& = &
\int_0^\epsilon d\beta\ sin(\beta) d_{K_i,K_i}^{J_i}(\beta) N(\beta) \nonumber \\
& + & 
\int_0^\epsilon d\beta'\ sin(\beta') d_{K_i,K_i}^{J_i}(\pi-\beta') N(\pi-\beta'), \nonumber \\
\end{eqnarray}
where $N(\beta)\equiv \langle \Phi \vert e^{i\beta J_y} 
\vert \Phi \rangle$, as in Eq. (43) of \cite{Villars}, and $\beta'=\pi-\beta$. 
In the first integral on the r.h.s., the approximation $\sin(\beta)\approx\beta$ holds to order $\beta^2$. Furthermore, the $d$-function can be approximated for $\beta\approx 0$ as~\cite{VMK}
\begin{eqnarray}\label{eq:d0}
d^{J}_{MM'}(\beta) & \approx & \frac{\xi_{MM'}}{\mu !}
\sqrt{\frac{(s+\mu+\nu)!(s+\mu)!}{s!(s+\nu)!}}\left(\frac{\beta}{2}
\right)^\mu \left[ 1 - \frac{2s(s+\mu+\nu+1)+\nu(\mu+1)}{2(\mu+1)} 
\left(\frac{\beta}{2}\right)^2 + \ldots \right], \nonumber 
\\
d^{J_i}_{K_iK_i}(\beta) & \approx & 1 - \left[ J(J+1) - K_i^2 \right] \left( \frac{\beta}{2} 
\right)^2,
\end{eqnarray}
where $\mu=\vert M-M'\vert = 0$, 
$\nu = \vert M+M'\vert = 2\vert K_i\vert$, $s = J - \frac{1}{2}
\left( \mu + \nu \right) = J - \vert K_i \vert$, $\xi_{MM'} = 1$ 
if $M=M'=K_i$ was used in the second line while truncating at order $\beta^2$. The formula is the same as Eq.~(38b) of Ref.~\cite{Villars}.
Similarly, for the second integral, one can obtain for 
$\beta\approx\pi$
\begin{eqnarray}\label{eq:d1}
d^{J}_{MM'}(\beta) & \approx & \frac{\xi_{MM'}}{\nu !}(-)^s
\sqrt{\frac{(s+\mu+\nu)!(s+\nu)!}{s!(s+\mu)!}}\left(\frac{\pi-\beta}{2}
\right)^\nu \nonumber \\
& \times & \left[ 1 - \frac{2s(s+\mu+\nu+1)+\mu(\nu+1)}{2(\nu+1)} 
\left(\frac{\pi-\beta}{2}\right)^2 + \ldots \right], \nonumber 
\\
d^{J_i}_{K_iK_i}(\beta) & \approx & \frac{1}{(2\vert K_i\vert)!}(-)^{J-\vert K_i\vert}
\frac{(J+\vert K_i\vert)!}{(J-\vert K_i\vert)!}\left(
\frac{\pi-\beta}{2}\right)^{2K_i} \nonumber \\
& = & (-)^{J-\vert K_i\vert} \left( \begin{array}{c} J+\vert K_i\vert 
\\ J-\vert K_i\vert \end{array} \right) \left(
\frac{\pi-\beta}{2}\right)^{2K_i},
\end{eqnarray}
which is the same as Eq.~(38d) in Ref.~\cite{Villars}. 
Finally, for $N(\beta)$, one can use the gaussian approximation~\cite{Villars}
\begin{equation}
N(\beta) \approx {\rm exp}\left( -\frac{\beta^2}{\beta_0^2} \right),
\end{equation}
where $\beta_0\equiv 2/\sqrt{\langle J_\perp^2 \rangle}$. 
Since $\langle J_\perp^2 \rangle$ is large in well-deformed nuclei displaying strong collective rotational motion,  the function $N(\beta)$ is strongly peaked as assumed at the start.
Equation (\ref{eq:appr1}) eventually becomes, to order $\beta$, 
\begin{eqnarray}
\int_{-1}^1 d(cos\beta)\ d_{K_i,K_i}^{J_i}(\beta) \langle \Phi_i \vert e^{i\beta J_y} 
\vert \Phi_i \rangle & = &
\int_0^\epsilon d\beta\ \beta\ e^{-\frac{\beta^2}{\beta_0^2}} \nonumber \\
& + & 
(-)^{J-\vert K_i \vert} \left( \begin{array}{c} J+\vert K_i\vert
\\ J-\vert K_i\vert \end{array} \right)
\int_0^{\epsilon} d\beta'\ \beta' \left( \frac{\pi-\beta}{2}\right)^{2K_i} N(\beta') 
\nonumber \\
& = & \left[ 1 + (-)^J \delta(K_i,0) \right] 
\int_0^\epsilon d\beta\ \beta\ e^{-\frac{\beta^2}{\beta_0^2}} \, .
\end{eqnarray}
In the last step, the necessity that $K_i=0$ for
the second contribution to be non-zero has been considered, and Eq.~(50a) of Ref.~\cite{Villars} stating that $N(\pi-\beta)=N(\beta)$ for $\beta \approx 0$.

Further assuming that $\epsilon$ is small enough for all the above approximations to be valid 
but large enough with respect to $\beta_0^2$, namely $\beta_0^2 \ll \epsilon^2 \ll 1$, then
\begin{eqnarray}
\int_0^\epsilon d\beta\ \beta\ e^{-\frac{\beta^2}{\beta_0^2}} & = & \frac{\beta_0^2}{2}, \\
\int_{-1}^1 d(cos\beta)\ d_{K_i,K_i}^{J_i}(\beta) \langle \Phi_i \vert e^{i\beta J_y}
\vert \Phi_i \rangle & = & \frac{\beta_0^2}{2} \left[ 1 + (-)^J \delta(K_i,0) \right], \\
N_i & = & \frac{2}{\beta_0} \left[ 1 + (-)^J \delta(K_i,0) \right]^{-\frac{1}{2}}.
\end{eqnarray}

Let us now consider Eq. (\ref{eq:start}). The dominant contributions to the integral, as above, come from either the integrand at $\beta \approx 0$ so that $| \Phi_2 \rangle$ and its
quantum number $K_2$ are unchanged, or at $\beta \approx \pi$ 
\begin{eqnarray}
&& \int_{-1}^1 d(cos\beta)\ d^{J_2}_{K_1-\mu,K_2}(\beta) \langle \Phi_1 \vert T_{\lambda\mu} 
e^{i\beta J_y} \vert \Phi_2 \rangle = \nonumber \\ 
&&
\int_{0}^\epsilon d\beta\ sin(\beta)\ d^{J_2}_{K_1-\mu,K_2}(\beta) 
\langle \Phi_{K_1} \vert T_{\lambda\mu} \vert \Phi_{K_2} \rangle  
N(\beta) + \nonumber \\
&& \int_{0}^\epsilon d\beta'\ sin(\beta')\ d^{J_2}_{K_1-\mu,K_2}(\pi-\beta') 
\langle \Phi_{K_1} \vert T_{\lambda\mu} \vert \Phi_{-K_2} \rangle N(\pi-\beta').
\end{eqnarray}
In the first term, $K_2+\mu=K_1$ and the $d$-function 
can be approximated as in Eq. (\ref{eq:d0}). In the second term, instead, one has $-K_2+\mu=K_1$. Therefore, 
the function $d^{J_2}_{-K_2,K_2}$ appears that can be approximated  using
the first line of Eq. (\ref{eq:d1}) using $\xi_{MM'}=1$, $\mu=2\vert K_2 
\vert$, $\nu=0$ and $s=J_2-\vert K_2 \vert$. Putting all together, 
\begin{eqnarray}
&& \int_{-1}^1 d(cos\beta)\ d^{J_2}_{K_1-\mu,K_2}(\beta) \langle \Phi_1 \vert T_{\lambda\mu} 
e^{i\beta J_y} \vert \Phi_2 \rangle = \nonumber \\ 
&&
\int_{0}^\epsilon d\beta\ \beta\ e^{-\frac{\beta^2}{\beta_0^2}}\   
\langle \Phi_{K_1} \vert T_{\lambda\mu} \vert \Phi_{K_2} \rangle  
+ \int_{0}^\epsilon d\beta'\ \beta'\ 
e^{-\frac{\beta^2}{\beta_0^2}}\ (-)^{J_2 - \vert K_2 \vert}\ 
\langle \Phi_{K_1} \vert T_{\lambda\mu} \vert \Phi_{-K_2} \rangle =
\nonumber \\
&& \frac{\beta_0^2}{2} \left[ 
\langle \Phi_{K_1} \vert T_{\lambda\mu} \vert \Phi_{K_2} \rangle +
(-)^{J_2 - K_2}\ 
\langle \Phi_{K_1} \vert T_{\lambda\mu} \vert \Phi_{-K_2} \rangle
\right]. 
\end{eqnarray}

Finally, one arrives at 
\begin{eqnarray}
\langle J_1 \vert \vert T_\lambda \vert \vert J_2 \rangle & = & \frac{(2J_1+1)(2J_2+1)}{2} N_1N_2 \sum_\mu
(-)^{J_1-K_1} \left( \begin{array}{ccc} J_1 & \lambda & J_2 \\ -K_1 & \mu & K_1-\mu \end{array}
\right) \nonumber \\
&& \times \int_{-1}^1 d(cos\beta)\ d^{J_2}_{K_1-\mu,K_2}(\beta) \langle \Phi_1 \vert T_{\lambda\mu} 
e^{i\beta J_y} \vert \Phi_2 \rangle \nonumber \\
& = & \frac{(2J_1+1)(2J_2+1)}{2} \frac{4}{\beta_0^2}
\left[ 1 + (-)^{J_1} \delta(K_1,0) \right]^{-\frac{1}{2}}
\left[ 1 + (-)^{J_2} \delta(K_2,0) \right]^{-\frac{1}{2}} \nonumber \\
&& \times \sum_\mu (-)^{J_1-K_1} \left( \begin{array}{ccc} J_1 & \lambda & J_2 \\ -K_1 & \mu & K_1-\mu \end{array}
\right) \frac{\beta_0^2}{2} \nonumber \\ 
&& \times \left[ 
\langle \Phi_{K_1} \vert T_{\lambda\mu} \vert \Phi_{K_2} \rangle +
(-)^{J_2 - K_2}\ 
\langle \Phi_{K_1} \vert T_{\lambda\mu} \vert \Phi_{-K_2} \rangle
\right] \nonumber \\
&& =
\left[ 1 + (-)^{J_1} \delta(K_1,0) \right]^{-\frac{1}{2}}
\left[ 1 + (-)^{J_2} \delta(K_2,0) \right]^{-\frac{1}{2}} 
(2J_1+1)(2J_2+1) \nonumber \\
&& \times 
\left[ (-)^{J_1-K_1}
\left( \begin{array}{ccc} J_1 & \lambda & J_2 \\ -K_1 & \mu & K_2 
\end{array} \right)
\langle \Phi_{K_1} \vert T_{\lambda\mu} \vert \Phi_{K_2} \rangle
+ \right. \nonumber \\
&& \left. 
(-)^{J_1-K_1}
\left( \begin{array}{ccc} J_1 & \lambda & J_2 \\ -K_1 & \mu & -K_2 
\end{array} \right)
(-)^{J_2 - K_2}\ 
\langle \Phi_{K_1} \vert T_{\lambda\mu} \vert \Phi_{-K_2} \rangle
\right] \, ,
\end{eqnarray}
\end{widetext}
which constitutes the so-called needle approximation. The same formula
was reported in the Appendix of Ref. \cite{PhysRevC.77.034317}, although with some typographic errors.
\vspace{0.5cm}

\bibliographystyle{apsrev4-2}
\bibliography{bibliography}

\end{document}